\documentclass[conference]{IEEEtran}
\IEEEoverridecommandlockouts
\usepackage{cite}
\usepackage{subcaption}
\usepackage{amsmath,amssymb,amsfonts}
\usepackage{algorithm}
\usepackage{algorithmic}
\usepackage{tikz}
\usepackage{graphicx}
\usepackage{textcomp}
\usepackage{xcolor}
\def\BibTeX{{\rm B\kern-.05em{\sc i\kern-.025em b}\kern-.08em
    T\kern-.1667em\lower.7ex\hbox{E}\kern-.125emX}}
\begin{document}

%\title{A Novel Image Encryption Scheme For Data Security and Privacy in IoT and Edge Networks Using Feature-Aware Segmentation and Chaotic Chain Transformations\\
\title{A Novel Feature-Aware Chaotic Image Encryption Scheme For Data Security and Privacy in IoT and Edge Networks\\
}

\author{\IEEEauthorblockN{Muhammad Shahbaz Khan\IEEEauthorrefmark{1}, Ahmed Al-Dubai\IEEEauthorrefmark{1}, Jawad Ahmad\IEEEauthorrefmark{2}, Nikolaos Pitropakis\IEEEauthorrefmark{1} and Baraq Ghaleb\IEEEauthorrefmark{1}}

\IEEEauthorblockA{\IEEEauthorrefmark{1}School of Computing, Engineering and the Built Environment, \\ Edinburgh Napier University, Edinburgh, UK.\\
Emails: \{muhammadshahbaz.khan, a.al-dubai, n.pitropakis, b.ghaleb\}@napier.ac.uk}

\IEEEauthorblockA{\IEEEauthorrefmark{2}Cyber Security Center, Prince Mohammad Bin Fahd University, Al-Khobar, Saudi Arabia\\
Email: jahmad@pmu.edu.sa}
}

%\and
%\and
%\and
%\and
%\and

\maketitle

\begin{abstract}
The security of image data in the Internet of Things (IoT) and edge networks is crucial due to the increasing deployment of intelligent systems for real-time decision-making. Traditional encryption algorithms such as AES and RSA are computationally expensive for resource-constrained IoT devices and ineffective for large-volume image data, leading to inefficiencies in privacy-preserving distributed learning applications. To address these concerns, this paper proposes a novel Feature-Aware Chaotic Image Encryption scheme that integrates Feature-Aware Pixel Segmentation (FAPS) with Chaotic Chain Permutation and Confusion mechanisms to enhance security while maintaining efficiency. The proposed scheme consists of three stages: (1) FAPS, which extracts and reorganizes pixels based on high and low edge intensity features for correlation disruption; (2) Chaotic Chain Permutation, which employs a logistic chaotic map with SHA-256-based dynamically updated keys for block-wise permutation; and (3) Chaotic chain Confusion, which utilises dynamically generated chaotic seed matrices for bitwise XOR operations. Extensive security and performance evaluations demonstrate that the proposed scheme significantly reduces pixel correlation---almost zero, achieves high entropy values close to 8, and resists differential cryptographic attacks. The optimum design of the proposed scheme makes it suitable for real-time deployment in resource-constrained environments.

\end{abstract}


\section{Introduction}

The rapid expansion of the Internet of Things (IoT) and edge computing has transformed various domains, including smart cities, industrial automation, and healthcare \cite{chataut2023unleashing, perwej2019internet, harmon2015smart}. These systems rely on distributed machine learning (ML) and artificial intelligence (AI) to process real-time data and make intelligent decisions on resource-constrained devices. However, ensuring the security and privacy of image data in such decentralized environments remains a significant challenge \cite{roman2013features}. Sensitive images, including medical scans, surveillance footage, and industrial monitoring data, are frequently transmitted and processed across heterogeneous networks, making them vulnerable to eavesdropping, unauthorized access, and adversarial attacks \cite{quantum, hathaliya2022adversarial, tang2024survey}. Traditional encryption techniques, such as AES and RSA, are not well-suited for IoT and edge-based AI applications due to their high computational complexity and inefficient handling of large image data \cite{srss, cellsecure}. Hence, lightweight and adaptive encryption algorithms are required to protect privacy while maintaining system scalability and efficiency.

\begin{figure*}[!t]
\centerline{\includegraphics[width=0.7\linewidth]{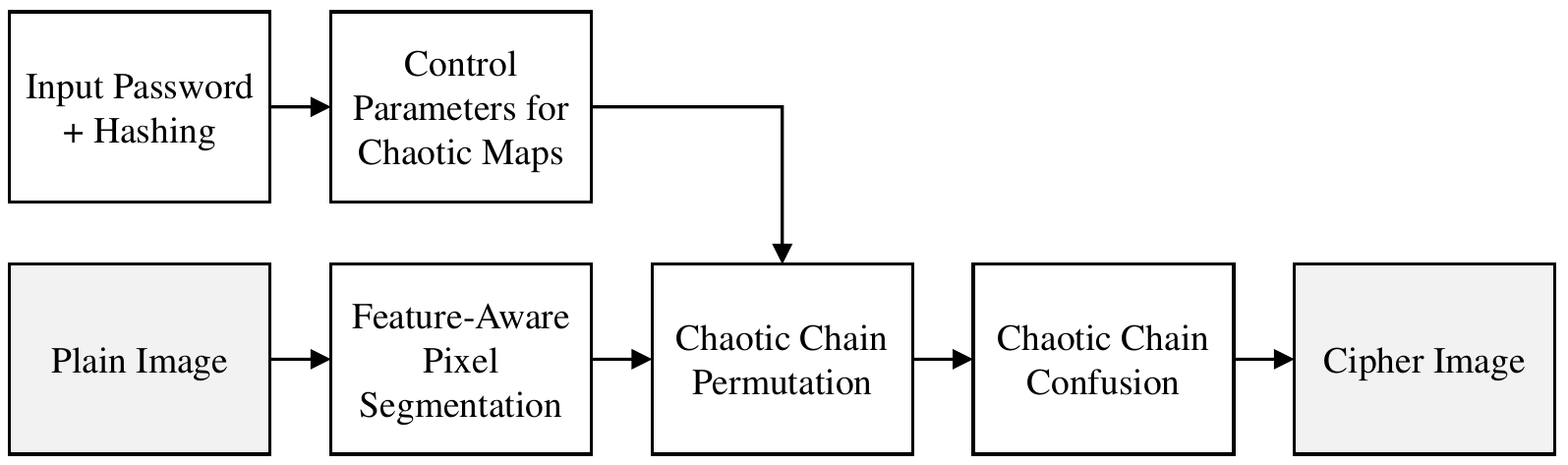}}
\caption{Overview of the Proposed Feature-Aware Encryption Scheme}
\label{fig1}
\end{figure*}

Chaos-based image encryption has emerged as a promising solution for securing image data in distributed environments \cite{umar2024chaos, permutex, lin2024image}. Chaotic systems possess properties such as sensitivity to initial conditions, pseudo-randomness, and ergodicity, making them well-suited for cryptographic applications\cite{zhang2024chaos, elkandoz2022image, KHAN20244214}. Various chaotic maps, including the Logistic map \cite{wu2022quantum}, Henon map \cite{chen2022hybrid}, and Lorenz system \cite{alexan2022rgb}, have been employed in image encryption schemes. These methods leverage permutation and substitution processes driven by chaotic sequences to disrupt pixel correlation and enhance security. However, existing chaotic encryption schemes often fail to meet the privacy and efficiency requirements of IoT and edge networks. Many approaches rely on static chaotic parameters, which can lead to periodic behaviour and reduced security. Additionally, conventional chaotic permutations are often independent of the underlying image structure, making them computationally inefficient for real-time ML-based edge analytics.

To address the aforementioned challenges, this paper proposes a privacy-preserving image encryption framework for IoT and edge-based intelligent systems, integrating Feature-Aware Pixel Segmentation (FAPS) with Chaotic Chain Permutation and Confusion mechanisms. By integrating these techniques, the proposed scheme enhances data protection and privacy. An overview of the proposed scheme is given in Fig. \ref{fig1}. The key contributions of this paper are as follows:
\begin{itemize}
    \item A novel \textit{Feature-Aware Pixel Segmentation} (FAPS) technique that optimizes image encryption for AI-driven IoT and edge networks by reducing correlation in image data. It extracts and reorganizes pixels based on high and low edge intensity features for effective correlation disruption.
    \item A \textit{ Chaotic Chain Permutation} method that employs a logistic chaotic map with SHA-256-based dynamic key generation, ensuring adaptive security and enhanced randomness.
    \item A \textit{Chaotic Chain Confusion} mechanism that utilises dynamically generated chaotic seed matrices for bitwise XOR operations in the confusion stage, making the encryption scheme resilient to cryptographic attacks.
\end{itemize}

The rest of this paper is structured as follows: Section II presents details on the proposed encryption scheme, including feature-aware pixel segmentation, chaotic chain permutation, and chaotic chain confusion. Section IV provides security analysis and experimental results, while Section V presents a clear and concise conclusion.

\section{The Proposed Feature-Aware Chaotic Image Encryption Scheme}

\begin{figure*}[!t]
\centerline{\includegraphics[width=0.99\linewidth]{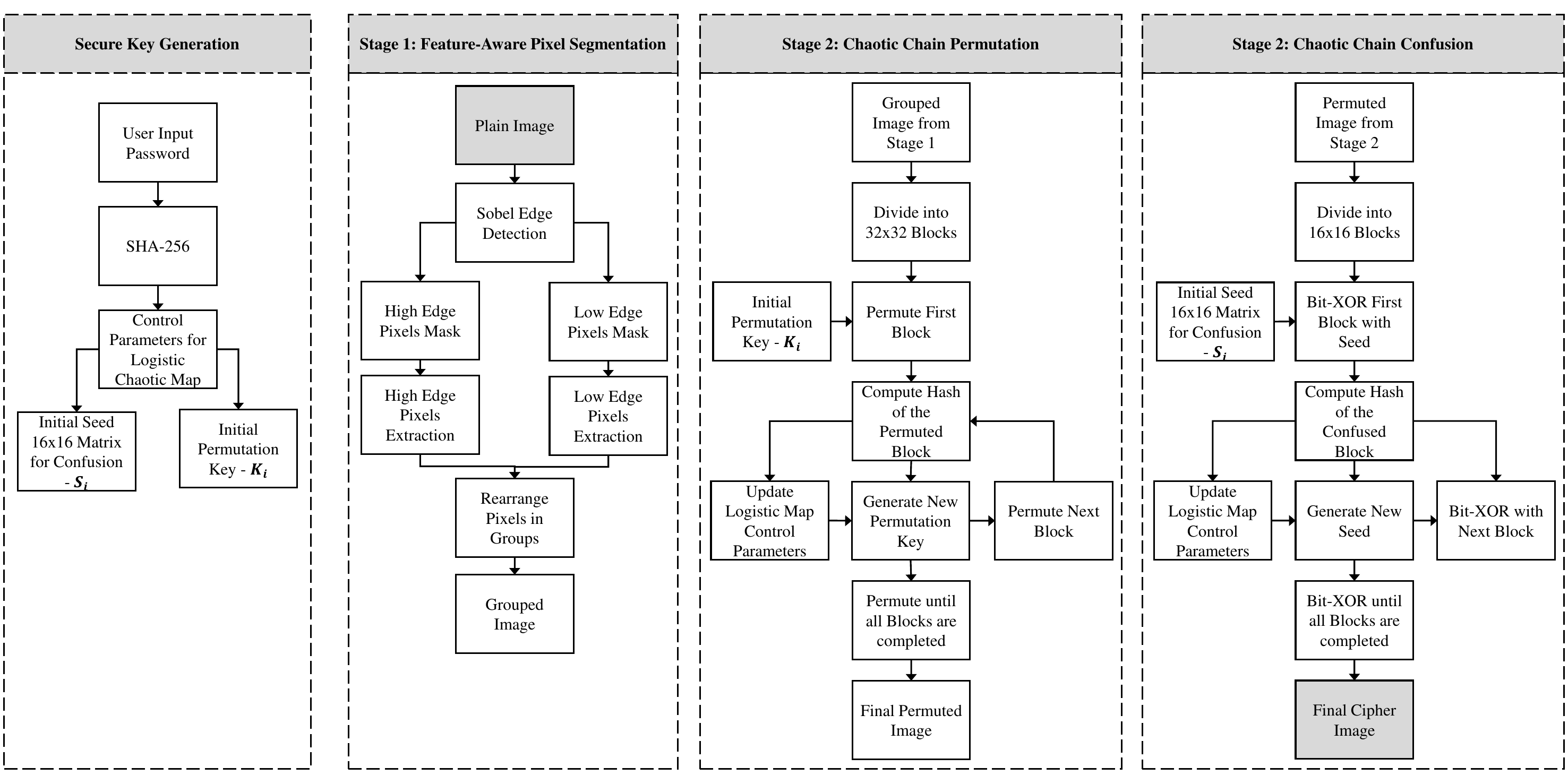}}
\caption{Complete block diagram of the proposed Feature-Aware Encryption Scheme}
\label{fig2:complete_scheme}
\end{figure*}

The proposed scheme consists of three stages: (1) \textit{Feature-Aware Pixel Segmentation}, which classifies pixels based on edge intensity to optimize encryption; (2) \textit{Chaotic Chain Permutation}, which applies a dynamically updated logistic chaotic map for block-wise permutation; and (3) \textit{Chaotic Chain Confusion}, which performs bitwise XOR with a dynamic chaotic seed matrices randomness. The complete block diagram of the proposed scheme entailing all three stages is given in Fig. \ref{fig2:complete_scheme} and are explained in the following subsections. In addition, a pseudo-code algorithm for the stepwise implementation of the proposed scheme is given in Algorithm \ref{alg:encryption}.

\subsection{Stage 1: Feature-Aware Pixel Segmentation (FAPS)}
\begin{figure*}[!t]
    \centering
    \begin{subfigure}{0.14\linewidth}
        \centering
        \includegraphics[width=.9\linewidth]{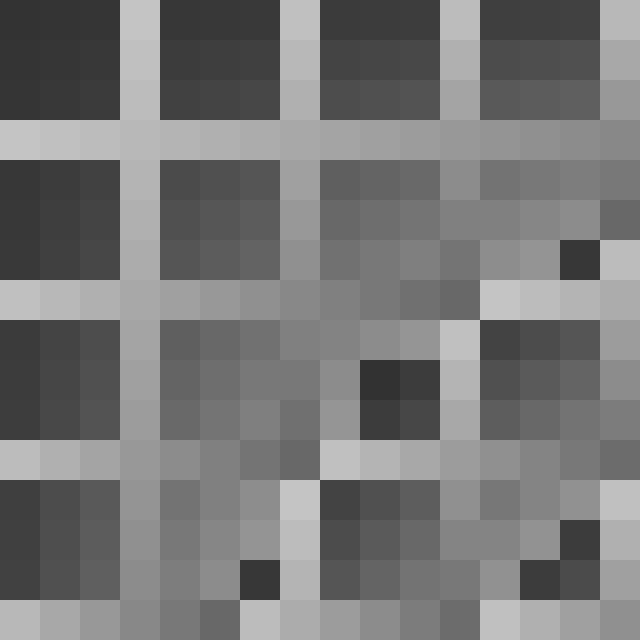}
        \caption{}
    \end{subfigure}
    \begin{subfigure}{0.14\linewidth}
        \centering
        \includegraphics[width=.9\linewidth]{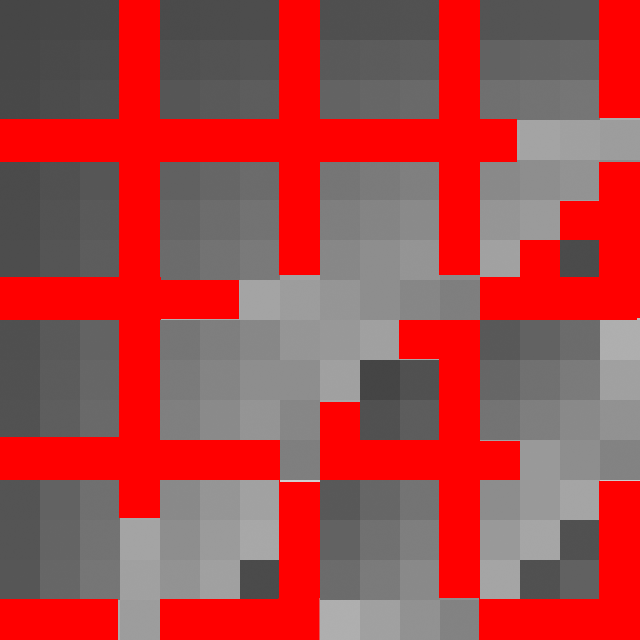}
        \caption{}
    \end{subfigure}
    \begin{subfigure}{0.14\linewidth}
        \centering
        \includegraphics[width=.9\linewidth]{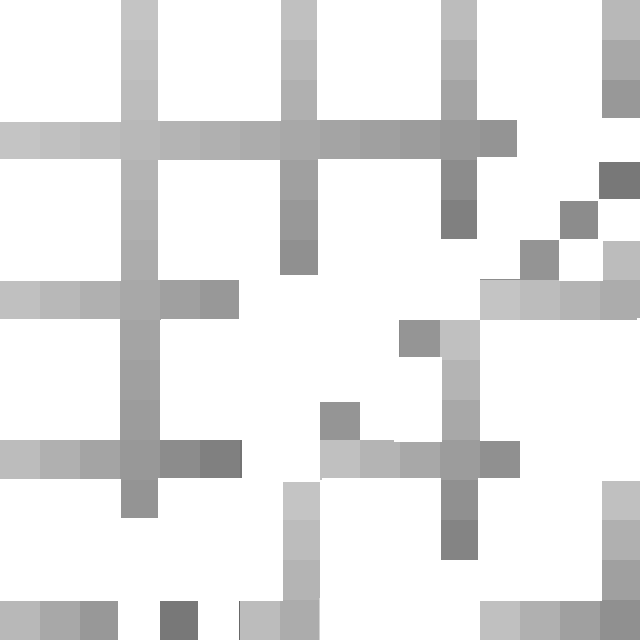}
        \caption{}
    \end{subfigure}
    \begin{subfigure}{0.14\linewidth}
        \centering
        \includegraphics[width=.9\linewidth]{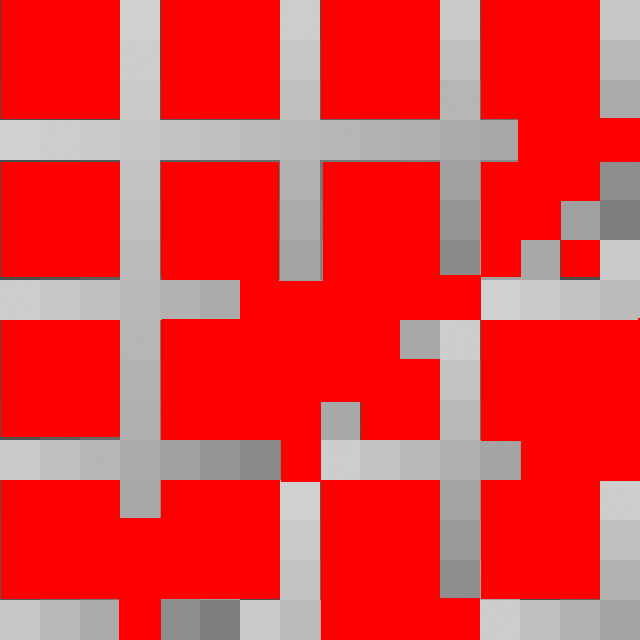}
        \caption{}
    \end{subfigure}
    \begin{subfigure}{0.14\linewidth}
        \centering
        \includegraphics[width=.9\linewidth]{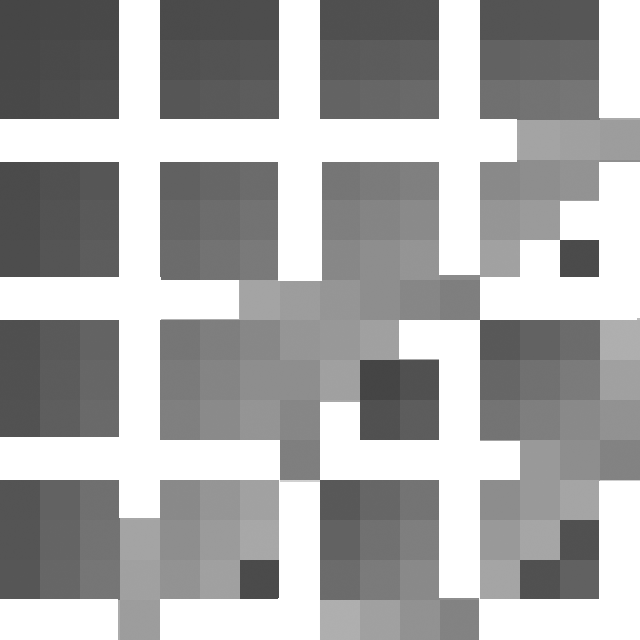}
        \caption{}
    \end{subfigure}
    \begin{subfigure}{0.14\linewidth}
        \centering
        \includegraphics[width=.9\linewidth]{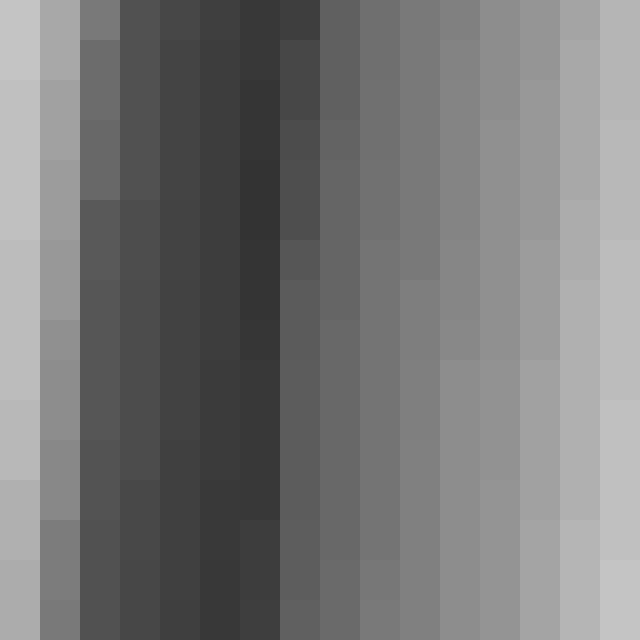}
        \caption{}
    \end{subfigure}
    \caption{Overview of the High Edge and Low Edge Pixel Classification in FAPS}
    \label{fig3:edge_detecttion}
\end{figure*}

\begin{figure}[!t]
    \centering
    \begin{subfigure}{0.48\linewidth}
        \centering
        \includegraphics[width=\linewidth]{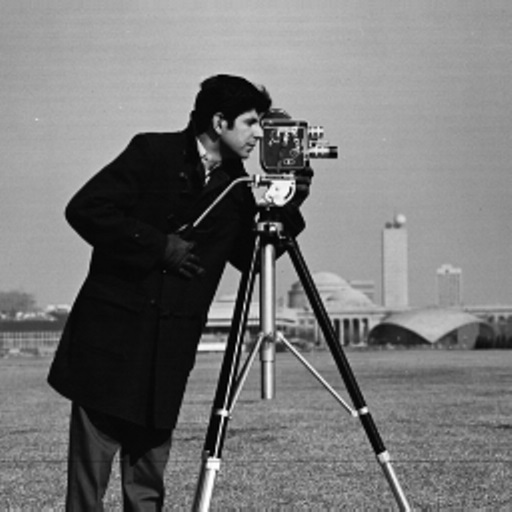}
        \caption{}
    \end{subfigure}
    \begin{subfigure}{0.48\linewidth}
        \centering
        \includegraphics[width=\linewidth]{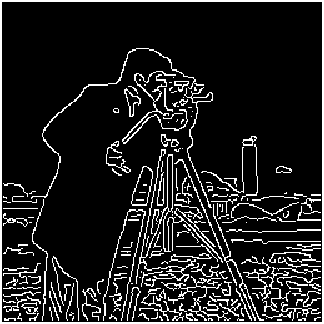}
        \caption{}
    \end{subfigure}
    
    \vspace{0.1cm} % Adds vertical spacing between rows

    \begin{subfigure}{0.48\linewidth}
        \centering
        \includegraphics[width=\linewidth]{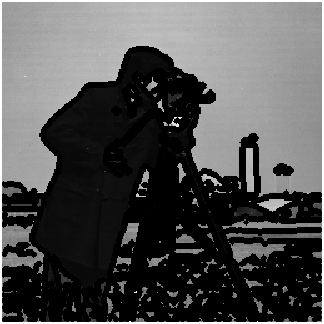}
        \caption{}
    \end{subfigure}
    \begin{subfigure}{0.48\linewidth}
        \centering
        \includegraphics[width=\linewidth]{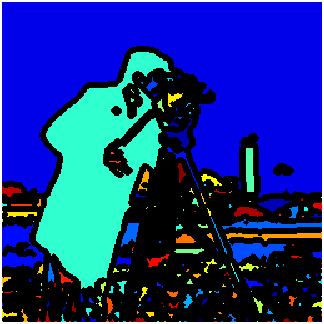}
        \caption{}
    \end{subfigure}
    
    \caption{Different stages of the feature extraction process on Cameraman image.}
    \label{fig4:cameraman_edge_detetction}
\end{figure}

This paper proposes a Feature-Aware Pixel Segmentation (FAPS) technique for preprocessing images before secure permutation. The method utilizes Sobel edge detection to segment pixels into high-variance and low-variance regions. The overview of variance classification for a $16 \times 16$ sample image is depicted in Fig. \ref{fig3:edge_detecttion}, whereas for a $256 \times 256$ Cameraman image, the process of edge detection with high and low variance region segmentation is depicted in Fig. \ref{fig4:cameraman_edge_detetction}.

Let $I(x,y)$ be the greyscale image of size $M \times N$, where each pixel has an intensity value in the range $I(x,y) \in [0,255]$. The proposed method follows these steps:

\subsubsection{Sobel Edge Detection}
The Sobel operator computes the gradient magnitude of each pixel to measure edge strength:
\begin{equation}
G_x = I(x,y) * S_x, \quad G_y = I(x,y) * S_y
\end{equation}
where $S_x$ and $S_y$ are the horizontal and vertical Sobel kernels:
\begin{equation}
S_x = \begin{bmatrix} -1 & 0 & +1 \\ -2 & 0 & +2 \\ -1 & 0 & +1 \end{bmatrix},
\quad
S_y = \begin{bmatrix} +1 & +2 & +1 \\ 0 & 0 & 0 \\ -1 & -2 & -1 \end{bmatrix}
\end{equation}

The edge magnitude at each pixel is then computed as:
\begin{equation}
G_{sobel}(x,y) = \sqrt{G_x^2 + G_y^2}
\end{equation}

The edge map is then normalized:
\begin{equation}
E(x,y) = \frac{G_{sobel}(x,y)}{\max(G_{sobel})}
\end{equation}
where $E(x,y) \in [0,1]$ represents the normalized edge intensity.

\subsubsection{High-Edge and Low-Edge Pixel Classification}
In this step,  a threshold $T$ is defined, which is obtained using Otsu’s method:
\begin{equation}
T = \arg\max_{\tau} [\sigma_B^2(\tau)]
\end{equation}
where $\sigma_B^2(\tau)$ is the between-class variance for a given threshold $\tau$. Using this threshold, we classify pixels into high-edge (HE) and low-edge (LE) regions:
\begin{equation}
P_{HE} = \{ I(x,y) \mid E(x,y) > T \}
\end{equation}
\begin{equation}
P_{LE} = \{ I(x,y) \mid E(x,y) \leq T \}
\end{equation}
where $P_{HE}$ contains textured and boundary regions, and $P_{LE}$ contains smooth regions.

\subsubsection{Pixel Sorting and Grouping}
To prepare for chaotic permutation, the pixels are reordered in a structured manner. High-edge pixels are sorted in descending order and placed in the upper half of the image:
\begin{equation}
P_{HE}' = \text{sort}(P_{HE}, \text{descend})
\end{equation}
On the other hand,  the low-edge pixels are sorted in ascending order and placed in the lower half:
\begin{equation}
P_{LE}' = \text{sort}(P_{LE}, \text{ascend})
\end{equation}

The final pre-permutation image $I'$ is defined as:
\begin{equation}
I' = \begin{bmatrix} P_{HE}' \\ P_{LE}' \end{bmatrix}
\end{equation}

\subsection{Stage 2: Chaotic Chain Permutation}
Once the image is preprocessed, a logistic map-based chaotic permutation is applied in a block-wise manner.

\begin{enumerate}
    \item The image $I'$ is divided into $B \times B$ non-overlapping blocks:
\begin{equation}
I' = \{ B_1, B_2, ..., B_k \}, \quad B_i \in \mathbb{R}^{b \times b}, \quad k = \frac{M \times N}{B^2}
\end{equation}
where each block $B_i$ has dimensions $32 \times 32$.

\item The logistic chaotic map is used to generate an initial permutation key and is defined as:
\begin{equation}
X_{n+1} = r X_n (1 - X_n)
\end{equation}
where $X_n \in (0,1)$ is the state variable, and $r$ is the chaotic control parameter. The initial key $X_0$ is chosen randomly within the chaotic range and is used to permute the first block.

\item A hash $H_1$ of the first permuted block $B_1$ is calculated using SHA-256. This hash is used to update the initial conditions and control parameters of the logistic map to generate a new permutation key for the 2nd block. This happens iteratively with each permuted block $B_i$. Each clock $B_i$ is permuted using a new permutation key and its hash value $H_i$ is computed using SHA-256 for the next block.
\begin{equation}
H_i = \text{SHA-256}(B_i')
\end{equation}
The chaotic system parameters are then updated:
\begin{equation}
X_0 = \frac{H_i}{2^{256}}, \quad r = 3.9 + 0.1 \times \left( \frac{H_i \mod 100}{100} \right)
\end{equation}

\item The process repeats for all blocks, ensuring that each block's permutation is influenced by the previous block's hash.

\item The permuted blocks are combined to form the final encrypted image:
\begin{equation}
I_{\text{perm}} = \begin{bmatrix} B_1' & B_2' & \dots & B_k' \end{bmatrix}
\end{equation}

\end{enumerate}

\subsection{Stage 3: Chaotic Chain Confusion}

After the permutation process, the image undergoes a block-wise confusion process using a logistic chaotic map and bitwise XOR operation to enhance security. This procedure ensures that each block is influenced by the previous block’s hash, making the confusion process highly dependent on initial conditions.

\begin{enumerate}
    \item The permuted image $I_{\text{perm}}$ is divided into $16 \times 16$ non-overlapping blocks:
    \begin{equation}
    I_{\text{perm}} = \{ B_1, B_2, ..., B_m \}, \quad B_i \in \mathbb{R}^{16 \times 16}, \quad m = \frac{M \times N}{16^2}
    \end{equation}
    where each block $B_i$ has dimensions $16 \times 16$.

    \item A chaotic seed matrix $S_1$ of size $16 \times 16$ is generated using the logistic map:
    \begin{equation}
    X_{n+1} = r X_n (1 - X_n)
    \end{equation}
    where $X_n \in (0,1)$ and $r$ is the chaotic control parameter. The initial seed matrix is given by:
    \begin{equation}
    S_1(i,j) = \lfloor 256 X_{n_{i,j}} \rfloor, \quad i,j = 1,2,\dots,16
    \end{equation}
    where $X_{n_{i,j}}$ are chaotic values mapped to integers in $[0,255]$.

    \item The first block $B_1$ is confused using a bitwise XOR operation with the seed matrix:
    \begin{equation}
    C_1 = B_1 \oplus S_1
    \end{equation}
    where $C_1$ is the confused output block.

    \item The confused block $C_1$ is hashed using SHA-256:
    \begin{equation}
    H_1 = \text{SHA-256}(C_1)
    \end{equation}
    This hash output is used to update the chaotic system parameters:
    \begin{equation}
    X_0 = \frac{H_1}{2^{256}}, \quad r = 3.9 + 0.1 \times \left( \frac{H_1 \mod 100}{100} \right)
    \end{equation}
    Using the updated parameters, a new chaotic seed matrix $S_2$ is generated:
    \begin{equation}
    S_2(i,j) = \lfloor 256 X_{n_{i,j}} \rfloor
    \end{equation}

    \item The process repeats iteratively for all blocks $B_i$, where each confused block $C_i$ influences the next seed matrix generation:
    \begin{equation}
    C_i = B_i \oplus S_i
    \end{equation}
    \begin{equation}
    H_i = \text{SHA-256}(C_i)
    \end{equation}
    \begin{equation}
    X_0 = \frac{H_i}{2^{256}}, \quad r = 3.9 + 0.1 \times \left( \frac{H_i \mod 100}{100} \right)
    \end{equation}
    where the updated values regenerate $S_{i+1}$ for the next block.

    \item After all blocks are processed, the final confused image $I_{\text{conf}}$ is obtained by combining all confused blocks:
    \begin{equation}
    I_{\text{conf}} = \begin{bmatrix} C_1 & C_2 & \dots & C_m \end{bmatrix}
    \end{equation}
\end{enumerate}

\begin{algorithm}[!b]
    \caption{Implementation of the Proposed Image Encryption Scheme}
    \label{alg:encryption}
    \begin{algorithmic}[1]
        \STATE \textbf{Input:} Grayscale image of size $M \times N$
        \STATE \textbf{Output:} Encrypted image
        
        \STATE \textbf{Step 1: Feature-Aware Pixel Segmentation}
        \STATE Apply Sobel edge detection to highlight texture and edges
        \STATE Compute edge map and normalize values
        \STATE Use Otsu’s method to classify high-texture and low-texture pixels
        \STATE Sort and group pixels based on feature classification
        \STATE Reconstruct the segmented image
        
        \STATE \textbf{Step 2: Chaotic Chain Permutation}
        \STATE Divide image into non-overlapping $32 \times 32$ blocks
        \STATE Initialize logistic chaotic map with an initial key
        \FOR{each block}
            \STATE Generate a unique permutation sequence using the chaotic map
            \STATE Permute block pixels according to the sequence
            \STATE Compute SHA-256 hash of the permuted block
            \STATE Update chaotic map parameters using the hash output
        \ENDFOR
        \STATE Reconstruct the permuted image
        
        \STATE \textbf{Step 3: Chaotic Chain Confusion}
        \STATE Divide image into non-overlapping $16 \times 16$ blocks
        \FOR{each block}
            \STATE Generate a dynamic chaotic seed matrix
            \STATE Perform bitwise XOR operation between the block and seed matrix
            \STATE Compute SHA-256 hash of the confused block
            \STATE Update chaotic map parameters using the hash output
        \ENDFOR
        \STATE Reconstruct the final encrypted image
    \end{algorithmic}
\end{algorithm}

\section{Results and Security Analysis}

This section evaluates the performance of the proposed encryption scheme through entropy analysis, correlation analysis, and differential attacks or sensitivity analysis to minor changes in the plaintext image.

\subsection{Histogram Analysis}
An effective encryption scheme should produce cipher images with a uniform histogram, ensuring resistance against frequency-based attacks. The histograms of the encrypted images as shown in Fig. \ref{fig5:encryption_results} demonstrate a near-uniform distribution of pixel intensities, indicating that the encryption process effectively diffuses pixel values.

\begin{table}[!b]
\caption{Correlation Evaluation}
\centering
\begin{tabular}{|c|p{1.5cm}|c|c|c|c|}
\hline
\textbf{Sr.} & \textbf{Test Image} & \textbf{Corr. Value} & \multicolumn{3}{c|}{\textbf{Correlation Coefficients}} \\ \cline{4-6} 
 & & & \textbf{Hor.} & \textbf{Ver.} & \textbf{Diag.} \\
\hline
1 & Cameraman & 0.00012 & -0.0006 & 0.0068 & -0.0071 \\
\hline
2 & Baboon & 0.00045 & 0.0028 & 0.0029 & 0.0021 \\
\hline
3 & Houses & 0.00038 & -0.0049 & -0.0036 & 0.0053 \\
\hline
\end{tabular}
\label{tab:correlation}
\end{table}

\begin{table}[!b]
\caption{Information Entropy Evaluation}
\centering
\begin{tabular}{|c|c|c|c|}
\hline
\textbf{Sr.} & \textbf{Image} & \textbf{Plain Image} & \textbf{Cipher Image} \\ \hline
1 & Cameraman & 7.448 & 7.998 \\ \hline
2 & Baboon & 7.051 & 7.998 \\ \hline
3 & Houses & 7.011 & 7.998 \\ \hline
\end{tabular}
\label{tab:entropy-analysis}
\end{table}

\subsection{Correlation Analysis}
\begin{figure}[!b]
    \centering
    % First Row
    \begin{subfigure}{0.25\linewidth}
        \centering
        \includegraphics[width=\linewidth]{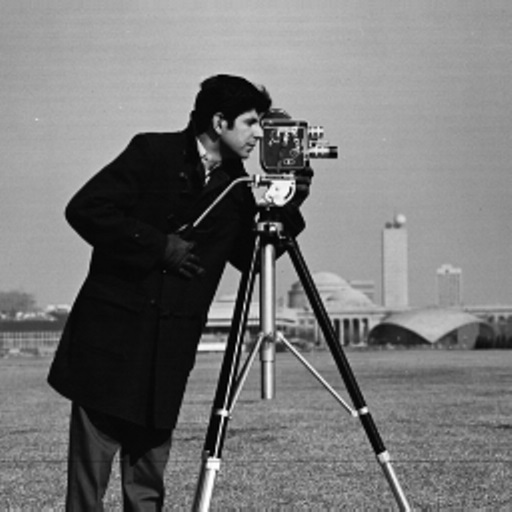}
        \caption{}
    \end{subfigure}
    \begin{subfigure}{0.25\linewidth}
        \centering
        \includegraphics[width=\linewidth]{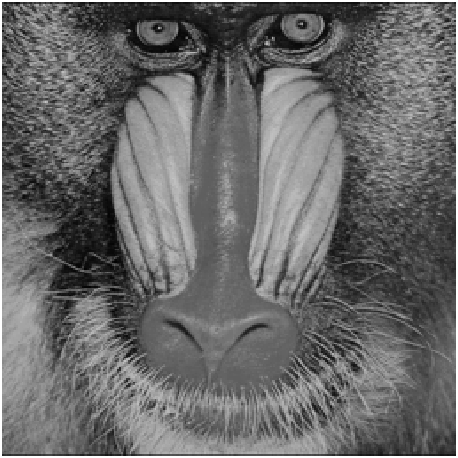}
        \caption{}
    \end{subfigure}
    \begin{subfigure}{0.25\linewidth}
        \centering
        \includegraphics[width=\linewidth]{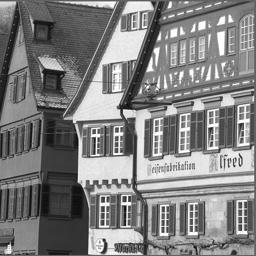}
        \caption{}
    \end{subfigure}

    \vspace{0.1cm} % Adds vertical spacing between rows

    % Second Row
    \begin{subfigure}{0.25\linewidth}
        \centering
        \includegraphics[width=\linewidth]{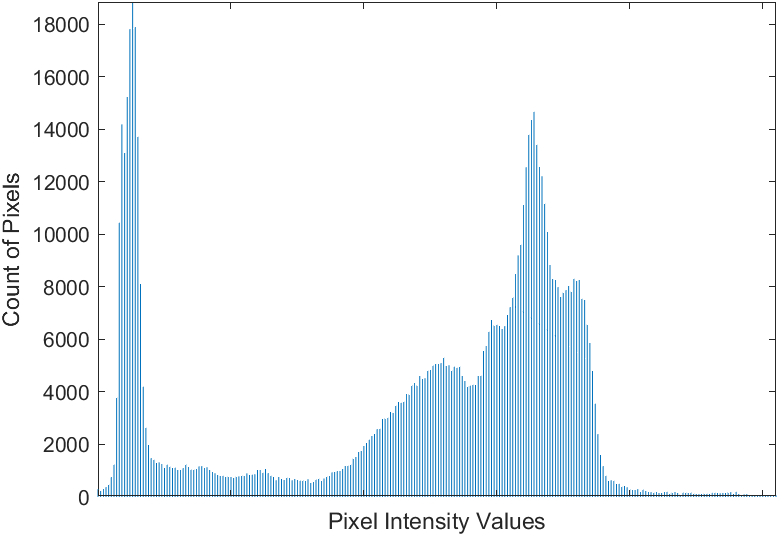}
        \caption{}
    \end{subfigure}
    \begin{subfigure}{0.25\linewidth}
        \centering
        \includegraphics[width=\linewidth]{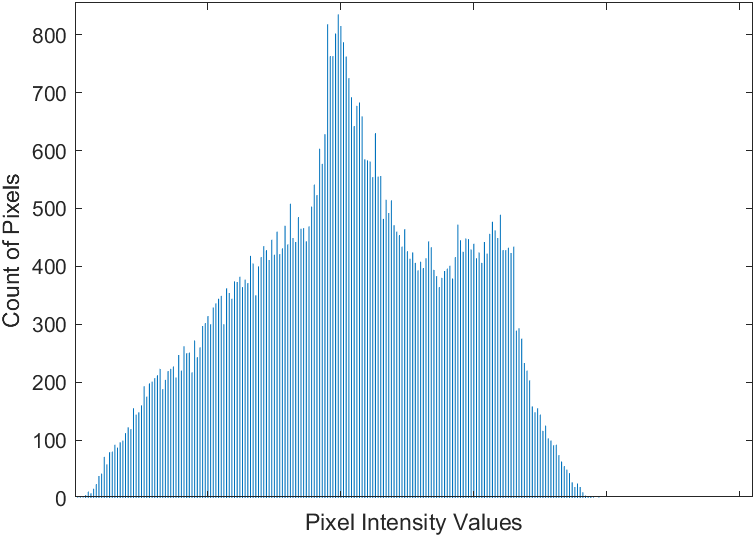}
        \caption{}
    \end{subfigure}
    \begin{subfigure}{0.25\linewidth}
        \centering
        \includegraphics[width=\linewidth]{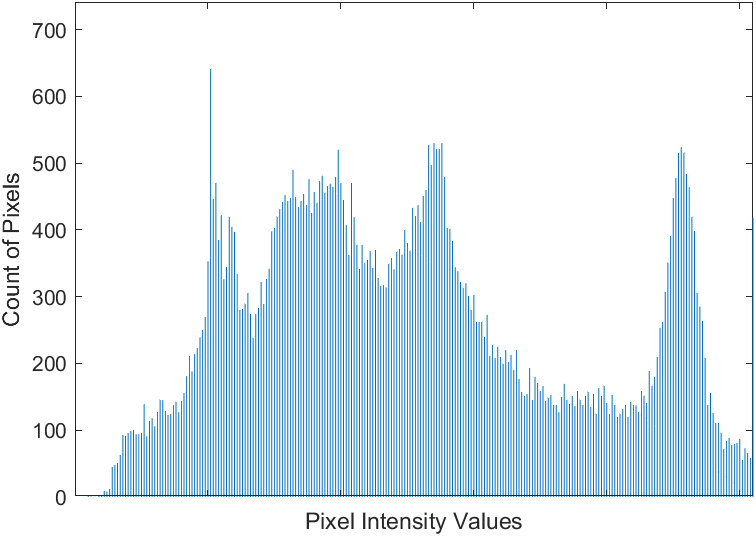}
        \caption{}
    \end{subfigure}

    \vspace{0.1cm} % Adds vertical spacing between rows

    % Third Row
    \begin{subfigure}{0.25\linewidth}
        \centering
        \includegraphics[width=\linewidth]{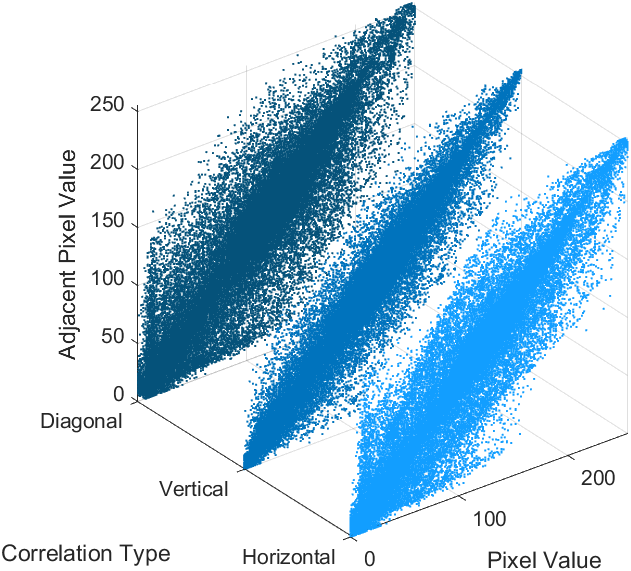}
        \caption{}
    \end{subfigure}
    \begin{subfigure}{0.25\linewidth}
        \centering
        \includegraphics[width=\linewidth]{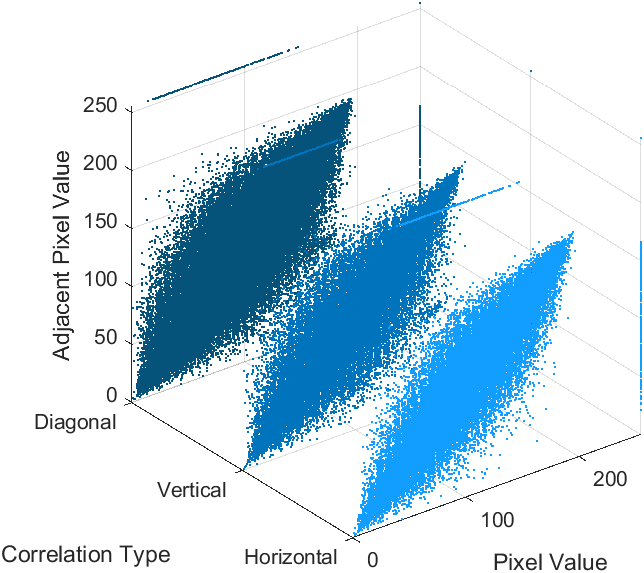}
        \caption{}
    \end{subfigure}
    \begin{subfigure}{0.25\linewidth}
        \centering
        \includegraphics[width=\linewidth]{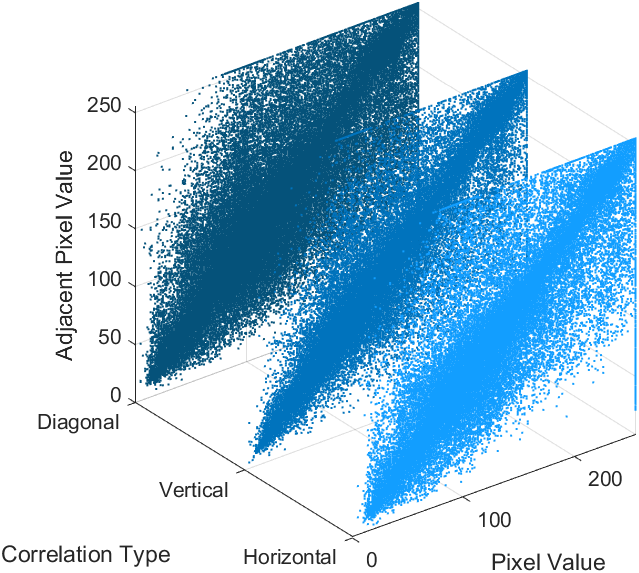}
        \caption{}
    \end{subfigure}

    \vspace{0.1cm} % Adds vertical spacing between rows

    % Fourth Row
    \begin{subfigure}{0.25\linewidth}
        \centering
        \includegraphics[width=\linewidth]{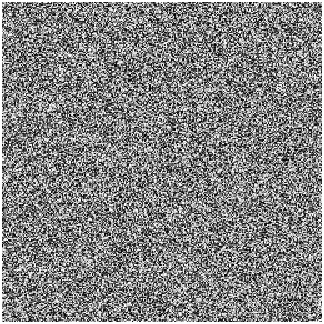}
        \caption{}
    \end{subfigure}
    \begin{subfigure}{0.25\linewidth}
        \centering
        \includegraphics[width=\linewidth]{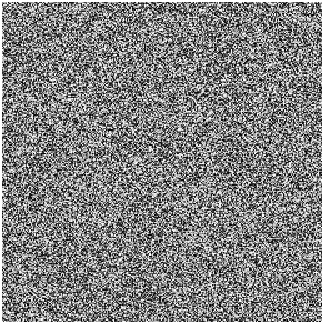}
        \caption{}
    \end{subfigure}
    \begin{subfigure}{0.25\linewidth}
        \centering
        \includegraphics[width=\linewidth]{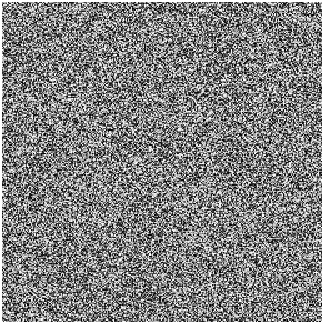}
        \caption{}
    \end{subfigure}

    \vspace{0.1cm} % Adds vertical spacing between rows

    % Fifth Row
    \begin{subfigure}{0.25\linewidth}
        \centering
        \includegraphics[width=\linewidth]{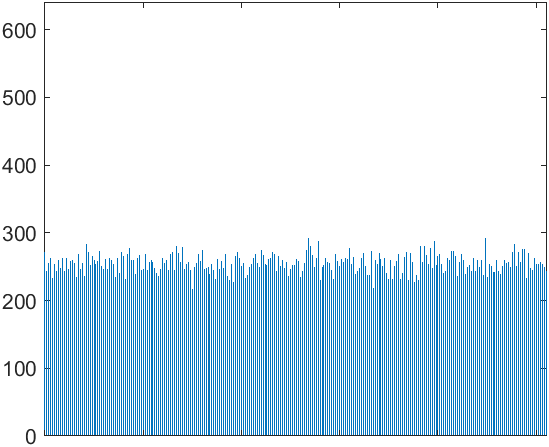}
        \caption{}
    \end{subfigure}
    \begin{subfigure}{0.25\linewidth}
        \centering
        \includegraphics[width=\linewidth]{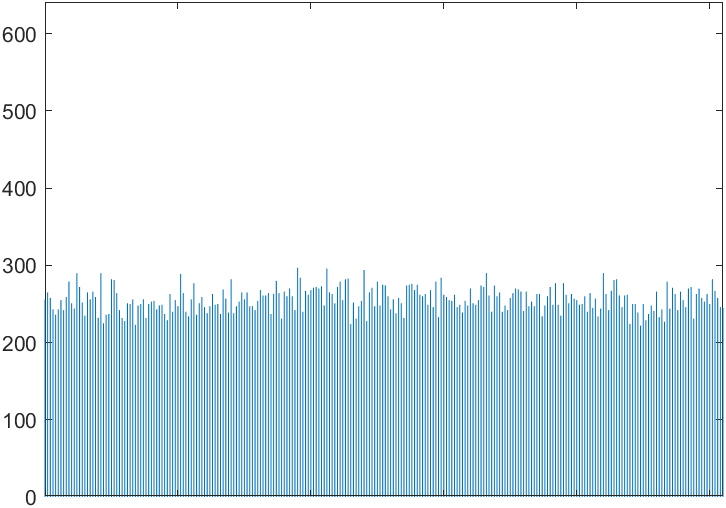}
        \caption{}
    \end{subfigure}
    \begin{subfigure}{0.25\linewidth}
        \centering
        \includegraphics[width=\linewidth]{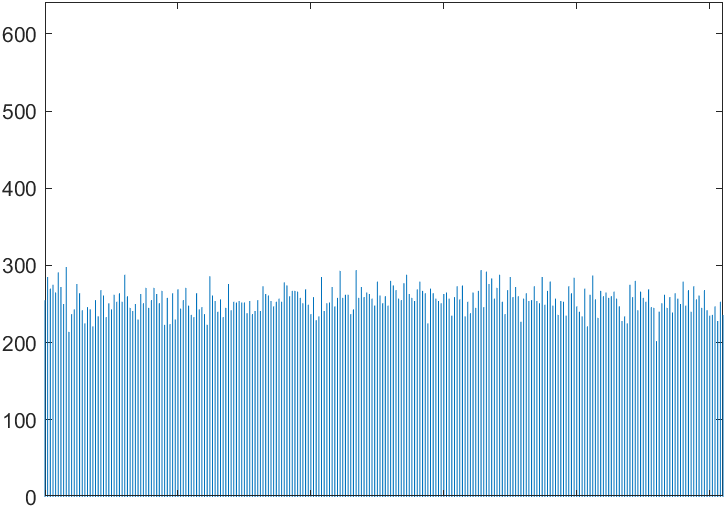}
        \caption{}
    \end{subfigure}

    \vspace{0.1cm} % Adds vertical spacing between rows

    % Sixth Row
    \begin{subfigure}{0.25\linewidth}
        \centering
        \includegraphics[width=\linewidth]{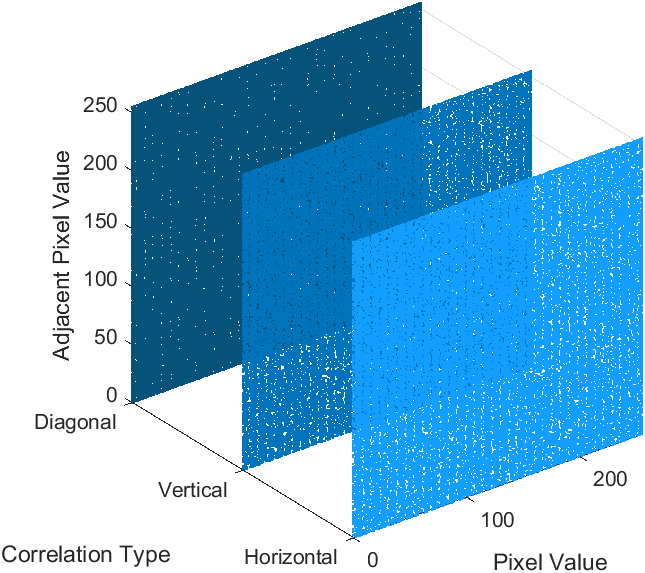}
        \caption{}
    \end{subfigure}
    \begin{subfigure}{0.25\linewidth}
        \centering
        \includegraphics[width=\linewidth]{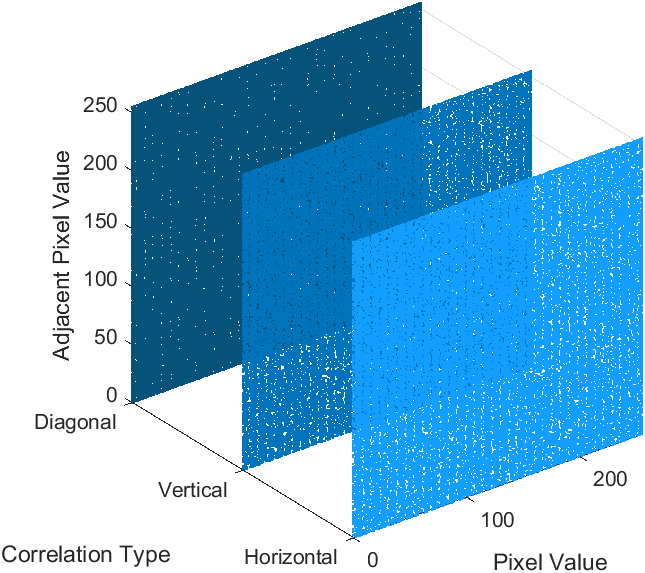}
        \caption{}
    \end{subfigure}
    \begin{subfigure}{0.25\linewidth}
        \centering
        \includegraphics[width=\linewidth]{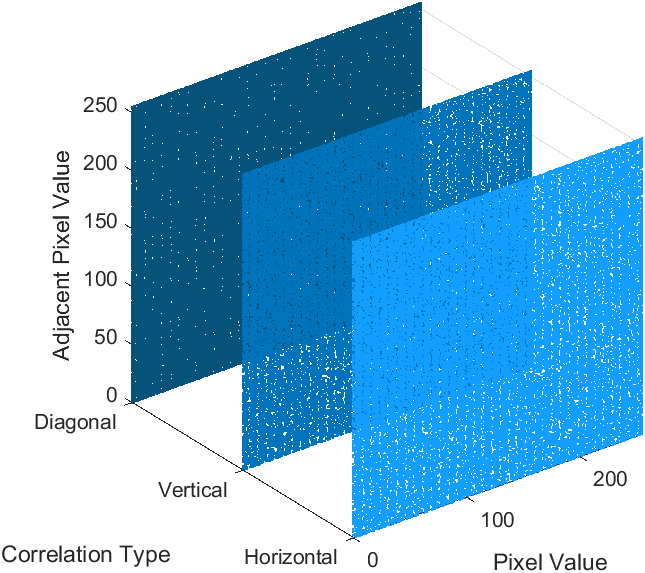}
        \caption{}
    \end{subfigure}

    \caption{Encryption results with histogram and correlation analysis}
    \label{fig5:encryption_results}
\end{figure}

Image encryption aims to eliminate pixel correlation to prevent statistical attacks. Table \ref{tab:correlation} shows the correlation coefficients for plain and cipher images. The plain images exhibit strong correlations due to natural redundancy, whereas the encrypted images achieve values close to zero in horizontal, vertical, and diagonal directions. Furthermore, Fig. \ref{fig5:encryption_results} depicts effective spread out of all correlation coefficients depicting maximum correlation disruption. The correlation is found by:

\begin{equation}
r = \frac{\sum_{i=1}^{N} (x_i - \mu_x)(y_i - \mu_y)}
         {\sqrt{\sum_{i=1}^{N} (x_i - \mu_x)^2} \sqrt{\sum_{i=1}^{N} (y_i - \mu_y)^2}}
\label{eq:correlation}
\end{equation}

where:
\begin{itemize}
    \item \( r \) is the correlation coefficient between adjacent pixels.
    \item \( x_i \) and \( y_i \) are the intensity values of two adjacent pixels.
    \item \( \mu_x \) and \( \mu_y \) are the mean intensity values of all pixels in the image.
    \item \( N \) is the total number of pixel pairs considered for correlation computation.
\end{itemize}

\subsection{Entropy Analysis}
Information entropy measures the randomness of an image, with an ideal entropy value for a perfectly encrypted image being close to 8. Table \ref{tab:entropy-analysis} presents the entropy values of plain images and their corresponding cipher images. The entropy of plain images is significantly lower due to redundant pixel structures, whereas the cipher images consistently achieve values near 7.998, indicating a highly unpredictable and secure encryption process. 

\begin{equation}
H(X) = - \sum_{i=0}^{255} P(x_i) \log_2 P(x_i)
\label{eq:entropy}
\end{equation}

where:
\begin{itemize}
    \item \( H(X) \) is the Shannon entropy of the image.
    \item \( P(x_i) \) represents the probability of occurrence of the intensity level \( x_i \).
    \item The summation runs over all possible intensity values from 0 to 255 for an 8-bit grayscale image.
\end{itemize}

\subsection{Differential Attack Resistance}
\begin{figure}[!t]
  \centering
  % First row
  \begin{subfigure}{0.25\linewidth}
      \centering
      \includegraphics[width=\textwidth]{01.jpg}
      \caption{}
      \label{fig:image1}
  \end{subfigure}
  \hfill
  \begin{subfigure}{0.25\linewidth}
      \centering
      \includegraphics[width=\textwidth]{02.png}
      \caption{}
      \label{fig:image2}
  \end{subfigure}
  \hfill
  \begin{subfigure}{0.25\linewidth}
      \centering
      \includegraphics[width=\textwidth]{03.jpg}
      \caption{}
      \label{fig:image3}
  \end{subfigure}

  % Second row
  \vspace{0.2em}
  \begin{subfigure}{0.25\linewidth}
      \centering
      \includegraphics[width=\textwidth]{01_enc.png}
      \caption{}
      \label{fig:image4}
  \end{subfigure}
  \hfill
  \begin{subfigure}{0.25\linewidth}
      \centering
      \includegraphics[width=\textwidth]{02_enc.png}
      \caption{}
      \label{fig:image5}
  \end{subfigure}
  \hfill
  \begin{subfigure}{0.25\linewidth}
      \centering
      \includegraphics[width=\textwidth]{03_enc.png}
      \caption{}
      \label{fig:image6}
  \end{subfigure}

  % Third row
  \vspace{.2em}
  \begin{subfigure}{0.25\linewidth}
      \centering
      \includegraphics[width=\textwidth]{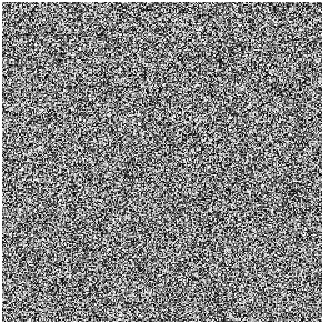}
      \caption{}
      \label{fig:image7}
  \end{subfigure}
  \hfill
  \begin{subfigure}{0.25\linewidth}
      \centering
      \includegraphics[width=\textwidth]{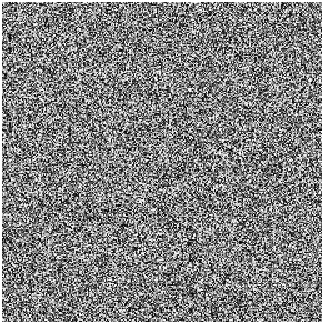}
      \caption{}
      \label{fig:image8}
  \end{subfigure}
  \hfill
  \begin{subfigure}{0.25\linewidth}
      \centering
      \includegraphics[width=\textwidth]{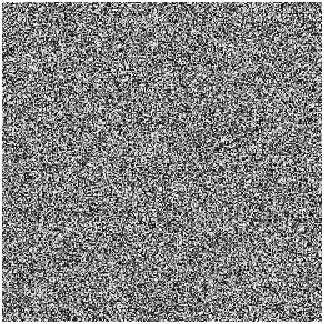}
      \caption{}
      \label{fig:image9}
  \end{subfigure}

  % Fourth row
  \vspace{0.2em}
  \begin{subfigure}{0.25\linewidth}
      \centering
      \includegraphics[width=\textwidth]{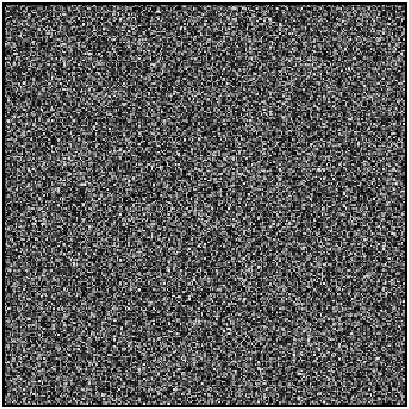}
      \caption{}
      \label{fig:image10}
  \end{subfigure}
  \hfill
  \begin{subfigure}{0.25\linewidth}
      \centering
      \includegraphics[width=\textwidth]{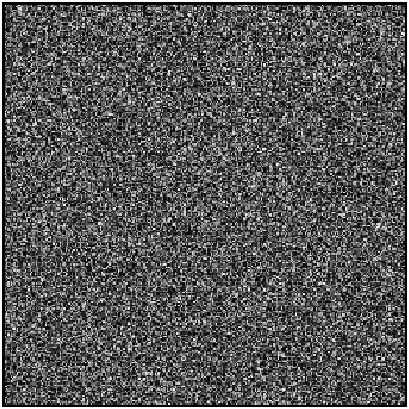}
      \caption{}
      \label{fig:image11}
  \end{subfigure}
  \hfill
  \begin{subfigure}{0.25\linewidth}
      \centering
      \includegraphics[width=\textwidth]{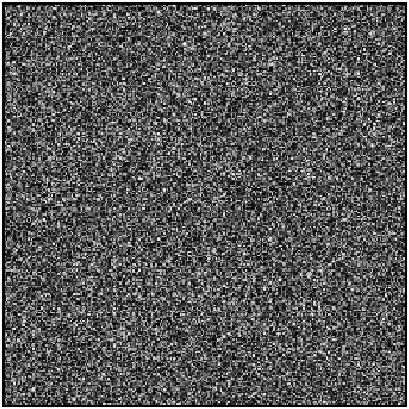}
      \caption{}
      \label{fig:image12}
  \end{subfigure}

  \caption{Differential attack/Sensitivity analysis. (a-c) plain test images. (d-f) original cipher images. (g-i) cipher images of one-bit corrupted plain images. (j-l) Difference between original and corrupted cipher images.}
  \label{fig6_differential_attacks}
\end{figure}

A highly secure encryption algorithm should be sensitive to even the smallest modifications in the plaintext. To evaluate this, a one-bit difference test was conducted, where an image was encrypted twice: first in its original form and then with a single-bit change in the plain image. The absolute difference between the two resulting cipher images was computed to analyse the propagation effect of the minor modification. The results in Fig. \ref{fig6_differential_attacks} demonstrate that the difference between the two encrypted images is substantial, highlighting the avalanche effect of the proposed encryption scheme.

\section{Conclusion}
This paper presented a novel feature-aware chaotic image encryption scheme designed to enhance security and privacy in IoT and edge networks. The proposed approach integrated Feature-Aware Pixel Segmentation, Chaotic Chain Permutation, and Chaotic Chain Confusion to effectively disrupt pixel correlation and improve resistance against statistical and differential attacks. Experimental results demonstrated that the scheme achieved near-ideal entropy values and significantly reduced correlation in encrypted images, ensuring strong security. Additionally, sensitivity analysis confirmed that the encryption process exhibited a high avalanche effect, making it resilient to differential attacks. The proposed method provided a lightweight yet robust encryption mechanism suitable for resource-constrained environments, thus contributing to secure image transmission and storage in intelligent distributed systems. Future work may explore hardware acceleration and adaptive chaotic models to further optimize performance and security.

% References section
\bibliographystyle{IEEEtran}
\bibliography{main}

\end{document}